\documentclass[preprint,prd,nofootinbib,tightenlines,amsmath]{revtex4}
\usepackage{graphicx}
\usepackage{bm}
\usepackage{slashbox}

\begin{document}
\baselineskip=15pt \parskip=5pt

\vspace*{3em}

\title{A Holographic Energy Model}

\author{Peng Huang$^{1}$}
\email{phuang@emails.bjut.edu.cn}
\author{Yong-Chang Huang$^1$$^{2}$$^3$}
\email{ychuang@bjut.edu.cn}
\affiliation{$^1$Institute of Theoretical Physics, Beijing University of Technology, Beijing 100124, China\\
$^2$Kavli Institute for Theoretical Physics, Chinese Academy of Sciences, Beijing 100080, China \\
$^3$CCAST (World Lab.), P.O. Box 8730, Beijing 100080, China}

\begin{abstract}
We suggest a holographic energy model in which
the energy coming from spatial curvature, matter and radiation can be obtained by using the particle horizon for
the infrared cut-off. We show the consistency between the
holographic dark-energy model and the holographic energy
model proposed in this paper. Then, we give a holographic
description of the universe.


\end{abstract}

\maketitle

\section{Introduction}

The holographic dark energy model \cite{1}- \cite{6} is a phenomenological model
which is simple and effective. Originally, Ref.\cite{1} suggested that,
in quantum field theory, due to the limit made by the formation of a
black hole, an ultraviolet (UV) cut-off is related to an infrared
(IR) cut-off. If $\rho _D $ is the quantum zero-point energy density
caused by a UV cut-off, the total energy in a region of size $L$
should not exceed the mass of a black hole of the same size, then,
$L^3\rho _D \le LM_p^2 $ \cite{1}, even general nonstationary black holes
are investigated \cite{7}\cite{88}. Thus one has $\rho _D =3C^2M_p^2 L^{-2}$,
here, $C$ is a numerical constant introduced for convenience and
$M_p $ is Planck mass \cite{2}. If one supposes that there is no
interaction between dark energy and matter, an inevitable result is
to use the future event horizon for infrared (IR) cut-off, only by
doing this can we deduce the correct equation of state (EOS) to
obtain an accelerated universe. The holographic dark energy model
developed from this viewpoint is as follows \cite{3}
\begin{equation}
\label{eq1}
\rho _D =3C^2M_p^2 L_E^{-2} ,
\end{equation}
where $C$ is a positive numerical parameter which is in favour of $C=1$ [2,
3], $M_p $ is the Plank mass, $L_E =a(t)r_E (t)$, the definition of $r_E
(t)$ is $\int\limits_0^{r_E (t)} {\frac{dr}{\sqrt {1-kr^2} }} =\frac{R_E
(t)}{a(t)}=\int\limits_t^\infty {\frac{dt}{a(t)}} $, and $R_E (t)$ is future
event horizon, $k$=1, 0, -1 corresponds to the closed, flat and open
universe, respectively.

The index of the EOS of dark energy derived from Eq.(\ref{eq1}) is
\cite{3}
\begin{equation}
\label{eq2}
\omega _D =-\frac{1}{3}(1+\frac{2}{C}\sqrt {\Omega _D } \cos \frac{\sqrt k
R_E (t)}{a(t)}),
\end{equation}
so, in the early universe, when $\Omega _D \to 0$, one has $\omega _{DE} \to
-\frac{1}{3}$; in the dark-energy-dominated era, $\Omega _D \to 1$ and
$\omega _D \to -1$, namely, dark energy evolves towards the cosmological
constant.

It is profound that a simple combination of the Planck scale and IR
cut-off $L_E$ gives an energy density comparable to the observed
dark energy. This can be understood in terms of the holographic
principle \cite{8} \cite{888}\cite{9}, saying that the area of any surface limits the
information content of adjacent spacetime regions at
$1.4\times10^{69}$ bits per square meter \cite{9}, which is thought to be
manifest in an underlying quantum theory of gravity. Such a basic
principle should have the property that it is universal and does not
hold only for special objects. Thus, since the dark energy has
already shown its holographic character, a natural generalization is
that the remnant kinds of energy in the universe should also have
their holographic characters which people don't know yet.

In this paper, we will show that the energy coming from spatial
curvature, matter and radiation together is holographic and appears
when we use the particle horizon for the IR cut-off, furthermore, we
give a holographic description of the universe.

The arrangement of the paper is as follows. In Sec.2 we suggest a holographic
energy model in which the energy coming from spatial curvature,
matter and radiation together can be obtained by using the particle
horizon to make the IR cut-off; in Sec.3 we study the consistency
between the two holographic energy models and give a holographic
description of the universe; the last section is the summary and
conclusion.

\section{Holographic energy from spatial curvature, matter
and radiation}

Assuming a $\Lambda$CDM model ($\Omega = 1$ ) , Seven-Year WMAP
Observations give the result
\begin{equation}
\label{eq3}
0.0133<\Omega_K<0.0084 (95\%CL),
\end{equation}
this limit weakens significantly if dark energy is allowed to
be dynamical, in fact a closed universe with a small positive curvature ($\Omega _K \sim 0.01)$ is compatible with observations,
\cite{10}-\cite{12}, we now do our investigating in the closed universe; for the
open universe with a negative spatial curvature the results can be
very similarly obtained from the investigations on those for the
closed universe.

In the closed universe, we know that
\begin{equation}
\label{eq3}
\rho _K \sim a(t)^{-2},
\quad
\rho _M \sim a(t)^{-3},
\quad
\rho _R \sim a(t)^{-4},
\end{equation}
where the subscripts $K$, $M$and $R$ are shortly denoted as the
spatial curvature, matter and radiation, respectively. And we
notice the character of $\rho \propto a^{-3(1+\omega )}$ that is obtained
from the Friedmann equation, from which we can get the corresponding
indexes of $\rho _K ,\;\rho _M ,\;\rho _R $ in their EOS as follows
\begin{equation}
\label{eq4}
\omega _K =-\frac{1}{3},
\quad
\omega _M =0,
\quad
\omega _R =\frac{1}{3}.
\end{equation}
Therefore, we can use the general energy density $\rho _{KMR} $ to
denote the sum of the three energy densities of (\ref{eq3}), then,
the index $\omega _{KMR} $ in the EOS of $\rho _{KMR} $ evolves from
$\frac{1}{3}$ in the early universe to $-\frac{1}{3}$ in the
dark-energy-dominated era, thus, when there exists a holographic
description for $E_{KMR} $ ($E_{KMR} $ is short for energy coming
from spatial curvature, matter and radiation together), the deduced
$\omega _{KMR} $ must satisfy this evolvement.

Similar to the research of establishing the holographic dark energy
model, $\rho _{KMR} $ also suffer the limitation made by the
formation of a black hole, and then an ultraviolet cut-off
corresponding to $\rho _{KMR} $ in a region of size $L$ is related
to an infrared cut-off corresponding to a black hole of the same
size, which is just
\begin{equation}
\label{eqcompliment} \rho _{KMR} =3C'^2M_p^2 L^{-2},
\end{equation}
where the similar parameter $C'$ is introduced for convenience.
Then, the task is to find a suited infrared cut-off $L$ which can
not only give the correct evolvement of $\omega _{KMR}$ but also
make the value of $\rho_{KMR} $ match the experimental data.

It has been pointed out that \cite{13}, if one uses the Hubble radius to
provide the IR cut-off and supposes that there is no interaction
between dark energy and matter, the deduced EOS is just for
pressureless matter. So the Hubble radius is not a good choice for
the holographic description of $E_{KMR} $. And the event horizon has
been used for the holographic description of dark energy, so, we now
investigate the particle horizon radius for the IR cut-off. In this
case, Eq.(\ref{eqcompliment}) turns to
\begin{equation}
\label{eq5} \rho _p =3C'^2M_p^2 L_p^{-2},
\end{equation}
the definition of $L_p $ is
\begin{equation}
\label{eq6}
L_p =a(t)r_p (t),
\end{equation}
$r_p (t)$ is defined by
\begin{equation}
\label{eq7}
\int\limits_0^{r_p (t)} {\frac{dr}{\sqrt {1-kr^2} }} =\frac{R_p
(t)}{a(t)}=\int\limits_0^t {\frac{dt}{a(t)}} ,
\end{equation}
or
\begin{equation}
\label{eq8}
\sqrt k r_p (t)=\sin \frac{\sqrt k R_p (t)}{a(t)}=\sin y_p ,
\end{equation}
where $k$=1, 0, -1 corresponds to closed, flat and open universe,
respectively, and $y_p =\frac{\sqrt k R_p (t)}{a(t)}$.

We can discover that the holographic energy density $\rho _p $ is just $\rho
_{KMR} $.

To see this, firstly, we derive the concrete form of the EOS of $\rho _p $
from Eq.(\ref{eq5}). Using the definitions $\Omega _p =\frac{\rho _p }{\rho _c }$
and $\rho _c =3M_p^2 H^2$, we can get from Eq.(\ref{eq5}) the following result
\begin{equation}
\label{eq9} HL_p =\frac{C'}{\sqrt {\Omega _p } }.
\end{equation}
Using Eq.(\ref{eq6}-\ref{eq9}) we obtain
\[
\mathop L\limits^. \\_p =HL_p +a\mathop r\limits^. \\_p (t)
\]
\[
=\frac{C'}{\sqrt {\Omega _p } }+a\cos y_p
\frac{d}{dt}\int\limits_0^t {\frac{dt}{a}}
\]
\begin{equation}
\label{eq10} =\frac{C'}{\sqrt {\Omega _p } }+\cos y_p ,
\end{equation}
so the rate of change of the $\rho _p $ with time is
\begin{equation}
\label{eq11} \frac{d\rho _p }{dt}=-6C'^2M_p^2 L_p^{-3} \mathop
L\limits^. \\_p =-2H\rho _p (1+\frac{1}{C'}\sqrt {\Omega _p } \cos
y_p ).
\end{equation}
On the other hand, because of the conservation of the energy--momentum
tensor, the evolution of the holographic energy density $\rho _p $ is
governed by
\begin{equation}
\label{eq12}
\frac{d}{da}(a^3\rho _p )=-3a^2p_p ,
\end{equation}
where $p_p $ denotes the pressure coming from $\rho _p $, thus we obtain $p_p
=-\frac{1}{3}\frac{d\rho _p }{d\ln a}-\rho _p $, and the EOS of the
holographic energy $\rho _p $ is characterized by the index
\begin{equation}
\label{eq13}
\omega _p =\frac{p_p }{\rho _p }=-\frac{1}{3\rho _p }\frac{d\rho _p }{d\ln
a}-1=-\frac{1}{3H\rho _p }\frac{d\rho _p }{dt}-1,
\end{equation}
inserting Eq.(\ref{eq11}) to Eq.(\ref{eq13}), finally, we get
\begin{equation}
\label{eq14} \omega _p =-\frac{1}{3}(1-\frac{2}{C'}\sqrt {\Omega _p
} \cos \frac{R_p }{a(t)}),
\end{equation}
where $k=1$ is taken.

We can see that when $C'=1$ then in the early universe $\Omega _p
\to 1$ because $\Omega _D \to 0$ at that time; when it is the
dark-energy-dominated era, $\Omega _p $ must be close to zero, we
thus can see from Eq.(\ref{eq14}) that $\omega _p $ evolves from
$\frac{1}{3}$ in the early universe to $-\frac{1}{3}$ in the
dark-energy-dominated era which is the same as $\omega _{KMR} $ (
see the discussion below Eq.(\ref{eq4})). But is it reasonable that
$C'=1$? In fact we can see $C'$ is also in favor of 1 if we notice
that the total energy in a region of size $L _p$ is
$\frac{4\pi}{3}L^3 _p\rho _{KMR}$, and the mass of a black hole of
the same size $L _p$ is $4\pi M^2 _p L _p$, in the extreme case we
can equate these two quantities, we can find that
\begin{equation}
\label{eqcompliment2}\rho _{KMR}=3M^2 _pL^{-2} _p
\end{equation}
which shows $C'=1$ and is consistent with the holographic dark
energy model.

Secondly, let's study the magnitude of $\rho _p $.

Huang and Li \cite{3} give a useful expression
\begin{equation}
\label{eq15} HL_E =\frac{C}{\sqrt {\Omega _D } }.
\end{equation}
Comparing Eq.(\ref{eq15}) with Eq.(\ref{eq9}) deduced in this paper,
we find that they have the same mathematical structure except for
$\Omega _p $ and $\Omega _D $, respectively. We know that $\Omega _D
=0.73\pm 0.04$ from the first year WMAP observations \cite{10}, when
$\rho _p $ is just $\rho _{KMR} $ and together with the fact that
$\rho _{KMR} $ now is dominanted by $\rho _M $ nowadays, we have
the result that $\Omega _p \approx \Omega _M =0.27\pm 0.04$ \cite{10},
inserting the values of $\Omega _p $ and $\Omega _D $ into
Eq.(\ref{eq15}) and Eq.(\ref{eq9}), respectively, we can derive that
$L_E $ and $L_p $ are at the same order of magnitude, and according
to Eq.(\ref{eq1}) and Eq.(\ref{eq5}) we can discover that the
magnitudes of $\rho _D $ and $\rho _p $ are at the same order of
magnitude. Thus, since the magnitude of $\rho _D $ matches the
experimental data well in the holographic dark energy model, the
magnitude of $\rho _p $ also matches the experimental data well if
we equate it to $\rho _{KMR} $.

We can also see this from another point of view. Noticing that the
particle horizon is comparable to the Hubble horizon nowadays, we
insert $L_p^{-1} \sim H_0 =1.51\times 10^{-42}Gev$ \cite{10} and $M_p
\approx 2.43\times 10^{18}Gev$ into Eq.(\ref{eq5}), we get the
result that $\rho _p \sim 10^{-47}Gev^4$, which is just at the same
order of the magnitude of $\rho _{KMR} $, so we again find that the
magnitude of $\rho _p $ matches experimental data well.

Therefore, since the evolvement of the EOS of $\rho _p $ is the same
as that for $\rho _{KMR} $ and the magnitude of $\rho _p $ matches
experimental data well if we equate it to $\rho _{KMR} $, we can now
see that the holographic energy density $\rho _p $ is just $\rho
_{KMR} $. So we have the conclusion that the energy comeing from
spatial curvature, matter and radiation together can be described by
the holographic $E_{KMR} $ whose energy density $\rho _p $ is given
by Eq.(\ref{eq5}), we then can change the index $p$ as $KMR$ for
comfortable, consequently, we have
\begin{equation}
\label{eq16} \rho _{KMR} =3C'^2M_p^2 L_p^{-2}
\end{equation}
with the parameter $C'$ in favor of 1.

Maybe someone has the conclusion that this holographic energy model
can only be established in a universe in which a particle horizon
can be found. The implying behind this is that this model will fail
when one faces universe models without particle horizon. In fact, it is well motivated that our universe does
not begin with a singular point but is a cosmological egg
with a little scale factor $a(0)$ ( $a( 0 )$ approaches zero but is
not equal to zero), i.e., a rather hot egg with very high temperature, density and curvature. Therefore, there must exist
a particle horizon, which can be seen from the definition $R_p = a\int \limits_0^t \frac{dt}{a(t)}$ of the radius of the particle horizon.

\section{The consistency between the two holographic energy models
and the holographic description of the universe}

From the discussion of the parameter $C'$ in the Sec.2, we know
that it is also in favor of $C'=1$ as that in the holographic dark
energy model\cite{2}\cite{3}, which shows the consistency between these two
holographic energy models. We here give furthermore discussions on
the consistency. From the result in the Sec.2, it is convenient to
set $C'$ equate to $C$, then, taking a derivative of the total
energy density of the universe with respect to time $t$, we have
\[
\frac{d}{dt}\rho _{total} =\frac{d}{dt}\rho _{DE} +\frac{d}{dt}\rho _{KMR}
\]
\begin{equation}
\label{eq17} =-6C^2M_p^2 L_E^{-3} \mathop L\limits^. \\_E -6C^2M_p^2
L_p^{-3} \mathop L\limits^. \\_p .
\end{equation}
And Ref.\cite{3} gives an expression
\begin{equation}
\label{eq18} \mathop L\limits^. \\_E =HL_E -\cos y_E .
\end{equation}
Inserting Eq.(\ref{eq10}) and Eq.(\ref{eq18}) into Eq.(\ref{eq17}), we obtain
\begin{equation}
\label{eq19}
\frac{d}{dt}\rho _{total} =\frac{-2\rho _{DE} }{L_E }(HL_E -\cos y_E
)+\frac{-2\rho _{KMR} }{L_p }(HL_p +\cos y_p ),
\end{equation}
using Eq.(\ref{eq9}) and Eq.(\ref{eq15}), Eq.(\ref{eq19}) can be rewritten as
\[
\frac{d}{dt}\rho _{total} =-2H\rho _{DE} (1-\frac{1}{C}\sqrt {\Omega
_{DE} } \cos y_E )-2H\rho _{KMR} (1+\frac{1}{C}\sqrt {\Omega _{KMR}
} \cos y_p )
\]
\begin{equation}
\label{eq20} =-2H\rho _{total} +\frac{2}{C}H(\frac{\rho
_{DE}^{\frac{3}{2}} }{\sqrt {\rho _c } }\cos y_E -\frac{\rho
_{KMR}^{\frac{3}{2}} }{\sqrt {\rho _c } }\cos y_p ).
\end{equation}
Because $\rho _c =\rho _D +\rho _M +\rho _R -\rho _K $ coming from one of
Friedmann eqations and $\rho _{total} =\rho _D +\rho _M +\rho _R +\rho _K
=\rho _D +\rho _{KMR} $, we can have
\[
\frac{d}{dt}\rho _{total} =-2H\rho _c -4H\rho _K
+\frac{2}{C}H(\frac{\rho _{DE}^{\frac{3}{2}} }{\sqrt {\rho _c }
}\cos y_E -\frac{\rho _{KMR}^{\frac{3}{2}} }{\sqrt {\rho _c } }\cos
y_p )
\]
\begin{equation}
\label{eq21} =-2H\rho _c [1+2\Omega _K -\frac{1}{C}\Omega
_{DE}^{\frac{3}{2}} \cos y_E +\frac{1}{C}\Omega _{KMR}
^{\frac{3}{2}}\cos y_p ],
\end{equation}
letting this equation equate zero to obtain the extremum point, we have
\begin{equation}
\label{eq22} 1+2\Omega _K -\frac{1}{C}\Omega _{DE}^{\frac{3}{2}}
\cos y_E +\frac{1}{C}\Omega _{KMR} ^{\frac{3}{2}}\cos y_p =0.
\end{equation}
One can know from the holographic dark energy model \cite{3} that the
expansion of the universe will never have a turning point so that
the universe will not re-collapse; the dark energy will dominate our
universe and $\Omega _{DE} \to 1^+$, thus, we can have $\Omega
_{KMR} \to 0^+$ and $\Omega _K \to 0^+$; furthermore, the universe
evolves towards a de Sitter universe, where $y_E
=\int\limits_t^\infty {\frac{dt}{a}} \propto \int\limits_t^\infty
{\frac{dt}{e^{Ht}}} =\frac{1}{H}e^{-Ht}$, when the dark energy
evolves towards the cosmological constant and when the considered
time is large enough, then $y_E \to 0$. Thus, we can see from above
discussions in this paragraph that $\Omega _{DE} =1$ is the solution
of Eq.(\ref{eq22}), which implies $C=1$.

On the other hand, because a lot of general physical processes
should satisfy quantitative causal relation with no-loss-no-gain
character \cite{14}\cite{15}, e.g., Ref.\cite{16} uses the no-loss-no-gain
homeomorphic map transformation satisfying the quantitative causal
relation to gain exact strain tensor formulas in Weitzenb\"{o}ck
manifold. In fact, some changes ( cause ) of some quantities in
Eq.(\ref{eq22}) must result in the relative some changes ( result )
of the other quantities in Eq.(\ref{eq22}) so that Eq.(\ref{eq22})'s
right side keep no-loss-no-gain, i.e., zero, namely, Eq.(\ref{eq22})
also satisfies the quantitative causal relation. Hence the
investigations are consistent.

So, when we consider the holographic $E_{KMR} $ model and the
holographic dark energy model simultaneously, we can see $C=1$; on
the other hand, when we only consider the holographic $E _{KMR}$
model, the parameter also have the same result. This shows the
consistency between the two holographic energy models. More
commonly, we know that the dark energy and $E_{KMR} $ can be
obtained by using respective horizon for their IR cut-off, this
correspondence between the energy and the horizon both in the
holographic dark energy model and in the holographic $E_{KMR} $
model also shows the consistency between the two models. This
consistency implies that the holographic descriptions of the
energies may be on the correct way to describe the universe. Based
on this consideration, we can now say that a closed physical
universe is holographic, it makes up of two holographic components:

(i) Holographic dark energy: $\rho _D =3C^2M_p^2 L_E^{-2} $;

(ii) Holographic $E_{KMR}$: $\quad \rho _{KMR} =3C^2M_p^2 L_P^{-2}
$.

When $t\to 0$, $R_p =\mathop {\lim }\limits_{t\to 0}
a(t)\int\limits_0^t {\frac{dt}{a(t)}} \to 0$, and $\rho _{KMR} \sim
R_p^{-2} \to \infty $, it corresponds to the big bang; when
$\int\limits_0^t {\frac{dt}{a(t)}} \sim \int\limits_t^\infty
{\frac{dt}{a(t)}} $, the magnitude of $\rho _D $ and $\rho _{KMR} $
is comparable to each other, the particle horizon and the event
horizon are both comparable to the Hubble horizon, and this is the
duration we stay at present; when $t$ is large enough, the universe
is dark energy dominated, so the universe looks like a de Sitter
universe that $H\sim \frac{\sqrt \Lambda }{M_p }$ and $a(t)\sim
e^{Ht}$, thus, $\rho _{DE} \sim M_p^2 (e^{Ht}\sin
^{-1}\int\limits_t^\infty {\frac{dt}{e^{Ht}})^{-2}} \sim \Lambda $
(from Eq.(\ref{eq1})) and $\rho _{KMR} \sim M_p^2 (e^{Ht}\sin
^{-1}\int\limits_0^t {\frac{dt}{e^{Ht}})^{-2}} \sim M_p^2 e^{-2Ht}$
(from Eq.(\ref{eq16})) during this time, namely, the dark energy
evolves towards the cosmological constant and the $E_{KMR} $ density
decays vary fast, however, still has a non-vanishing value which is
proportional to $e^{-2Ht}$.

A deduction from the holographic description of the universe is that there
must be both dark energy and $E_{KMR} $ as long as the two horizons exist in
a given closed universe. Thus, for example, a closed de Sitter universe with
only a positive cosmological constant, in which the two horizons appear, can
not exist as a real physical universe but be a good approximation for the
real physical universe during dark energy dominated era since the $E_{KMR} $
decays so fast.

\section{Summary and Conclusion}

The motivation to study the holographic characteristic of the
energy coming from spatial curvature, matter and radiation
is that the holographic principle is believed to be a basic
principle which must be manifest in an underlying quantum
theory of gravity, and such a basic principle should have the property of universality and does not holds only for special object, thus, since the dark energy has already shown
its holographic character, a natural generalization is that the
remnant kinds of energy in the universe should also have
their holographic characters. In the holographic dark-energy
model \cite{2}\cite{3}, using an event horizon for the IR cut-off is an
inevitable choice, only by doing this can we get the correct
equation of state to accelerate the expansion of the universe.
In order to give a holographic model of the remnant energy
in the universe, similarly, we must find a suitable IR cut-off
that can give the correct equation of the state for these remnant energy, we can find that the particle horizon is also an
inevitable choice.

It is well known that the early universe is radiation-dominanted,
and the energy coming from spatial curvature decays slower than
those from radiation and matter when the universe is expanding,
which can be seen from Eq.(\ref{eq3}), so the index $\omega $ of the
EOS of the general energy density $\rho _{KMR} $, which denotes the
energy $E_{KMR} $ density coming from spatial curvature, matter and
radiation, must evolve from $\frac{1}{3}$ in the early time of the
universe to $-\frac{1}{3}$ when the universe is
dark-energy-dominanted. Similar to using the event horizon for the
IR cut-off in the holographic dark energy model, we then investigate
the particle horizon for the IR cut-off, we denote the new
holographic energy density got by this way as $\rho _p $, we find
that the index $\omega $ of the deduced EOS of $\rho _p $ shows the
expected behavior as that for $\rho _{KMR} $, and the magnitude of
$\rho _p $ also matches experimental data well, and we have the
conclusion that $E_{KMR} $ can be obtained by using the particle
horizon for the IR cut-off, and then we have established the
holographic $E_{KMR} $ model.

Furthermore, we study the consistency between the two holographic
energy models. We show that the both holographic models have
consistent requirements for the parameter $C$, and in the both
models the relative correspondences between the energy and the
horizon naturally shows their consistency. This consistency implies
that the holographic description of the energy is on the correct way
to describe the universe. Motivating by this point of view, we
propose the holographic description of the universe. According to
this description, a closed accelerated physical universe is
holographic and made up by two holographic components: the
holographic dark energy and the holographic $E_{KMR} $, the
evolution of the universe depends on the evolution of the two
components. The novel natures of this paper are not only that we
first suggest a holographic energy model, in which the energy coming
from spatial curvature, matter and radiation can be obtained by
using the particle horizon for the infrared cut-off, but also that a
holographic description of the universe is obtained, according to
the description, there must be both the holographic dark energy and
the holographic $E_{KMR} $ in the universe with the particle horizon
and the event horizon according to the holographic description of
the universe, so we argue the de Sitter universe, which has both the
two horizons.can not exist as a real physical universe but be a good
approximation for the real physical universe during dark energy
dominated era since the $E_{KMR}$ decays so fast.

We want to highlight that different components of the
observed energy density are associated with different holographic screens. The dark energy is associated with one
screen (the event horizon) which is presented in Eq.(\ref{eq1}) while the
remnant energy densities are associated with another screen
(the particle horizon) given in Eq.(\ref{eq16}) seriously in this paper.
It needs to be pointed out that, in general, FRW universe models
don't have simultaneously a particle horizon and a event horizon,
they may have one or other but not both at the same time. There is
no future event horizon in the decelerated universe, there is also
no particle horizon in the accelerated universe. However, the
particle horizon can always be found in any universe once we take a
short cutoff in the definition. Furthermore, the percentage of the
$\rho_{KMR}$ in the total energy density $\rho_t$ is decreasing
while the expansion of the universe is proceeding, and the
$\rho_{KMR}$-dominated epoch is turning to the dark-energy-dominated
epoch, which denotes our universe changes from decelerating to
accelerating, thus, in this picture, the universe have the both
horizons at the same time.

For a lot of further investigations, it is valuable to investigate,
e.g., the all relative investigations about the holographic $E_{KMR}
$, which are similar to those about the holographic dark energy in
different models, and so on.

\end{document}